# Role of surface states and band modulations in ultrathin ruthenium interconnects

Gyungho Maeng [a], Subeen Lim [a], Mi Gyoung Lee [b,c], Bonggeun Shong [d], Kyeongjae Cho [e], Yeonghun Lee [a,c,*]

[a]Department of Electronics Engineering, Incheon National University, Yeonsu-gu, Incheon 22012, Republic of Korea
[b]Department of Materials Science and Engineering, Incheon National University, Incheon 22012, Republic of Korea
[c]Research Institute for Engineering and Technology, Incheon National University, Yeonsu-gu, Incheon 22012, Republic of Korea
[d]Major in Advanced Materials and Semiconductor Engineering, Hanyang University, Ansan-si, Gyeonggi-do 15588, South Korea
[e]Department of Materials Science and Engineering, The University of Texas at Dallas, Richardson, TX 75080, USA

**ABSTRACT** Mitigating the RC delay from transistor miniaturization is essential for next-generation devices, driving a focus on interconnect electrical performance. Current copper-based interconnects face a critical challenge, that their resistivity sharply increases at the nanometer-scale due to surface and grain boundary scattering. Therefore, there is a pressing need for techniques that reduce resistance in ultrathin metal films. In this study, we employ the density functional theory to investigate how the intrinsic electronic structure of thin films impacts conductivity as a function of thickness. Notably, our analysis of ruthenium slab structures shows that surface states significantly influence thickness-dependent resistivity. It reveals that vacuum-terminated Ru slab exhibits decreasing resistivity with the decrease in thickness, whereas oxygen-terminated Ru slab shows the opposite trend. This difference is fundamentally attributed to the presence or absence of surface states, highlighting the importance of surface engineering in optimizing interconnect performance.

## 1. Introduction

Following the invention of the first transistor at Bell Laboratory in 1947[1], continuous efforts have been made to increase the integration density of integrated circuits[2–4]. As process nodes advance into the sub-nanometer regime, the physical space available for interconnects inevitably shrinks. Since the electrical resistance of a metal line is inversely proportional to its cross-sectional area, this reduction in dimensions leads to a drastic increase in resistance. Furthermore, the scaling of device features narrows the spacing between adjacent interconnects, thereby increasing the parasitic capacitance. Consequently, in contrast to transistors, which typically gain speed and energy efficiency with miniaturization, interconnects suffer from severe performance degradation. This performance limit is primarily governed by the RC delay, which has become the dominant factor restricting the operation speed of modern integrated circuits[5–9]. Copper (Cu), the industry-standard material adopted through dual-damascene processing, now faces a critical challenge as the metal pitch shrinks below 30 nm[10,11]. The primary issue is a sharp increase in resistivity due to enhanced surface and grain boundary scatterings at nanometer-scale thicknesses. Furthermore, the non-scalable barrier and liner layers, required to prevent Cu diffusion and enhance reliability, significantly increase the resistance of interconnects because the thickness of barriers and liners cannot be reduced below critical dimensions without degrading interconnect reliability[12,13]. To overcome this fundamental limitation, extensive research is underway to identify promising alternatives by theoretically classifying their intrinsic electrical properties[14,15]. Among them, ruthenium (Ru) is known to be one of the promising candidates for next-generation interconnects[16–18]. It exhibits lower resistivity than Cu at thin thicknesses. Furthermore, Ru enables stable and uniform nucleation on $SiO_2$ allowing for its use in barrierless interconnect schemes and requires only a thin adhesion layer to prevent delamination during chemical mechanical polishing[19,20].

Although the electrical resistivity of bulk metals is theoretically well established[21–24], predicting it at the nanoscale remains a challenge. Semi-classical models, such as the Fuchs-Sondheimer and Mayadas-Shatzkes models, have been used to account for increased scattering at the surface and grain boundaries in nanoscale[25–27]. However, these models become limited as they overlook how surface termination or quantum confinement alters the electronic band structure. Specifically, the presence of surfaces modifies the band structure, creating additional electronic states, known as surface states, that are distinct from those in the bulk[28]. These states can form a surface electron gas, potentially enhancing conductivity[29–31]. As dimensions shrink, the contribution of these surface conduction paths becomes increasingly important.

*Corresponding author.
E-mail address: y.lee@inu.ac.kr (Y. Lee).

In this context, we employ the first-principles calculations to simulate the thickness-dependent resistivity of Ru thin films. Specifically, we investigate the contrasting behaviors of vacuum-terminated and oxygen-terminated Ru slabs to elucidate how the presence or absence of surface states affects transport properties. As a noble metal with resistance to oxidation, Ru is likely to better preserve surface states, in contrast to Cu, which readily oxidizes under ambient conditions[32–34]. This study offers new insights into the design of low-resistivity nanoscale interconnects through surface-state engineering.

## 2. Computational Detail

The first-principles calculations were performed using the open-source software package Quantum ESPRESSO v.7.2[35,36], employing DFT with a plane-wave basis set. For the exchange-correlation term, the generalized gradient approximation (GGA) was implemented through the Perdew-Burke-Ernzerhof (PBE) functional[37,38]. The GGA-PBE functional was chosen as it has been demonstrated to accurately describe a range of physical properties of hexagonal close-packed (HCP) Ru—such as lattice parameters, bulk modulus, heat capacity, and thermal expansion—showing good agreement with experimental data[39]. Scalar-relativistic projector augmented-wave pseudopotentials were adopted[40–43]. Spin polarization was considered in all calculations. The kinetic energy cutoff for the plane-wave expansion of the wavefunctions was set to 52 Ry to ensure sufficient convergence of the total energy. For the charge density and potential, a higher cutoff of 353 Ry was applied to accurately describe the augmentation charges intrinsic to the PAW method. The bulk Ru primitive cell was calculated using an $18 \times 18 \times 12$ Monkhorst-Pack **k**-point grid, and the optimized HCP lattice constants were determined to be $a = 2.718$ Å and $c = 4.290$ Å[44–46]. All slab structure optimizations were performed using a $15 \times 15 \times 1$ **k**-point grid. While the unit cell lattice parameters were fixed to the optimized bulk values, the positions of all atoms in the entire slab were fully allowed to relax until the forces on each atom were below 0.001 Ry/Bohr. To exclude any artifacts such as spurious electric field, we analyzed the planar-averaged electrostatic potential along the surface normal. This analysis confirmed the absence of any built-in potential or dipole moment across the slab (detailed potentials are shown in Fig. S2). To analyze the thickness-dependent transport properties, electronic structures were calculated using a dense $150 \times 150 \times 6$ **k**-point grid (see Fig. S1 for the convergence test).

The transport properties based on semi-classical transport were evaluated using the BoltzTraP2[47,48], which generates denser **k**-point mesh by interpolating the original band structures derived from the DFT. The conductivity tensor is given by:

$$\sigma_{\alpha\beta} = \frac{e^2}{8\pi^3} \sum_n \int_{BZ} d^3k \, \tau_n(\mathbf{k}) v_{\alpha,n}(\mathbf{k}) v_{\beta,n}(\mathbf{k}) \frac{df_{FD}(E)}{dE} \bigg|_{E=E_n(\mathbf{k})} \quad (1)$$

where, $\alpha$ and $\beta$ denote the tensor indices, $e$ is the electron charge, BZ indicates the integration over the Brillouin zone, and $n$ is a band index. The carrier relaxation time is denoted by $\tau_n(\mathbf{k})$, $v_{\alpha,n}(\mathbf{k})$ is group velocity for each tensor direction, and the derivative of the Fermi-Dirac distribution function $\frac{df_{FD}(E)}{dE}$ is the Fermi window. Under the constant relaxation time approximation, $\tau_n(\mathbf{k})$ is assumed to be independent of the band indices and wave vectors, i.e., $\tau_n(\mathbf{k}) = \tau$. This simplifies the conductivity expression to a form proportional to $\tau$:

$$\sigma_{\alpha\beta} = \frac{e^2 \tau}{8\pi^3} \sum_n \int_{BZ} d^3k \, v_{\alpha,n}(\mathbf{k}) v_{\beta,n}(\mathbf{k}) \frac{df_{FD}(E)}{dE} \bigg|_{E=E_n(\mathbf{k})} \quad (2)$$

Here, the electrical resistivity, defined as the inverse of conductivity, was evaluated by combining the first-principles calculation results with experimental data. To address the inherent electrical anisotropy of HCP Ru, an effective isotropic value was derived by averaging the diagonal components $\sigma_{xx}$, $\sigma_{yy}$, and $\sigma_{zz}$. From our calculation, the product of the bulk resistivity and relaxation time ($\rho_0 \times \tau$) was determined to be $6.47 \times 10^{-22}$ Ω·m·s. This value was then divided by the experimental bulk resistivity of 7.80 μΩ·cm to extract the $\tau$[49,50]. This procedure yielded $\tau = 8.30$ fs for bulk Ru at 300 K, which was then used to calculate the slab resistivity under the constant relaxation time approximation.

*Corresponding author.
*E-mail address:* y.lee@inu.ac.kr (Y. Lee).

## 3. Results and discussion

We investigated room-temperature electrical properties of Ru thin films as a function of thickness and different surface terminations using the first-principles calculation. Bulk HCP Ru was first optimized to obtain lattice parameters $a = b = 2.718$ Å and $c = 4.290$ Å, which were used for supercell and slab structures. The films were modeled as $1 \times 1$ slabs derived from bulk Ru, cleaving along the thermodynamically most stable (0001) plane (Fig. 1)[51,52]. A vacuum layer of 15 Å was introduced to create slab models. To distinguish the effect of surface states, we compared two different systems. The first was a bare, vacuum-terminated Ru slab (Fig. 1(b)) and the second was an oxygen-terminated Ru-O slab (Fig. 1(c)) (details of atomic structures are shown in section 2 of the Supplementary Information). We constructed the Ru-O slab with a full 1 monolayer of oxygen coverage, fully passivating both surfaces to suppress the intrinsic surface states characteristic. For this configuration, oxygen atoms occupy the energetically favorable HCP-hollow sites on both surfaces of the Ru slab[44,53]. The calculated average Ru-O bond distance in the optimized structure is 1.98 Å. This value agrees well with the experimentally measured value, $2.05 \pm 0.16$ Å[53].

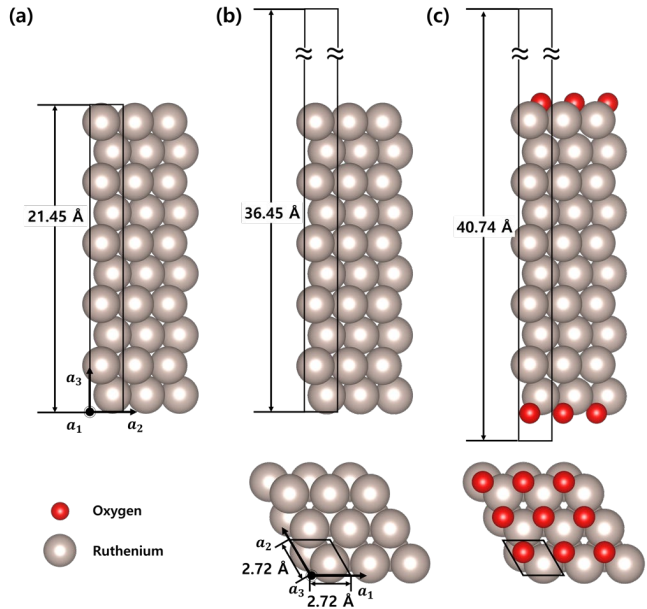

**Fig. 1.** Atomic structures of a Ru supercell and slabs consisting of 10 Ru layers. (a) Side view of the Ru supercell. (b) Side and top views of the Ru slab. (c) Side and top views of the Ru-O slab with oxygen atoms adsorbed on surfaces. The detailed atomic positions of both slabs are included in the supplementary section 2.

To elucidate the effect of surface conditions on the electronic structure of the Ru slab, we calculated the Fermi surfaces and band structures for both the Ru slab and the Ru-O slab models. Fig. 2(a) displays the Fermi surface of Ru. To provide a reference for the slab model analysis, the Fermi surface of a bulk Ru supercell is shown in Fig. 2(b). Since the Fermi surface of Ru is complex due to the overlapping $d$-bands, we employ projected Fermi surfaces to identify surface states. In contrast, for simpler metals such as Cu, surface states can be directly visualized through their wavefunctions[31]. The projected Fermi surfaces for the Ru and Ru-O slabs are illustrated in Fig. 2(c) and Fig. 2(d), respectively. A series of all projections are weighted by the orbital contributions from Ru atoms on both surface layers, with the color intensity representing the magnitude of the contribution. A comparison of the Fermi surfaces (Fig. 2(c) and Fig. 2(d)) reveals a key difference: the Ru slab exhibits significantly higher surface contributions near the Fermi level compared to the Ru-O slab, indicating the presence of additional electronic states at the surface of the Ru slab. These two slab models effectively define theoretical upper and lower bounds for the electrical transport properties of Ru thin films. The bare Ru slab, maintaining its intrinsic conductive surface states, represents an optimal case with maximal surface state contribution. Conversely, the fully oxygen-passivated Ru-O slab, where surface states are suppressed due to strong orbital hybridization, illustrates a scenario with minimal surface state contribution. In realistic interconnect environments, where interfacial conditions such as partial oxidation or various capping/barrier layers are present, the actual transport properties are expected to fall between these two curves, contingent upon the extent to which the conductive surface states are preserved or modified by the interfacial environment.

*Corresponding author.
*E-mail address:* y.lee@inu.ac.kr (Y. Lee).

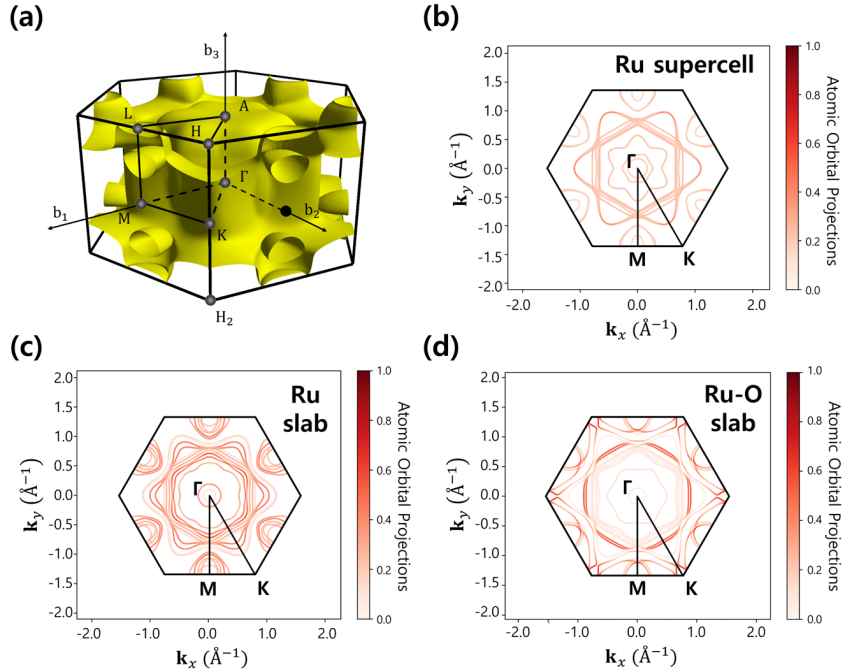

**Fig. 2.** (a) Fermi surface of the bulk Ru primitive cell and high-symmetry points in HCP Brillouin zone. Projected 2D Fermi surfaces of the (b) bulk Ru supercell, (c) Ru slab, and (d) Ru-O slab with 10 Ru layers displayed in Fig. 1. All projections (b-d) are weighted by the orbital contributions from Ru atoms on both surface layers, with the color intensity representing the magnitude of the contribution.

To further understand the nature of these surface states, we examine the band structures and projected density of states (PDOS) of the Ru and Ru-O slabs, illustrated in Fig. 3. The color intensity in the band structures represents the magnitude of the contribution from both surface Ru atoms. As shown in Fig. 3(a), the band structure of Ru slab reveals a strong band projection near the Fermi level ($E_F$), particularly between the Γ and M points. This is corroborated by the PDOS in Fig. 3(b), which shows that the states near $E_F$ are dominated by Ru $d$-orbitals (blue line). Since a high density of states generally implies a large number of available charge carriers, the substantial population of these states suggests conditions favorable for high electrical conduction. Further, a direct comparison with the bulk Ru DOS (Fig. S3(a)) reveals that the PDOS of the surface Ru atoms (Fig. 3(b)) exhibits higher density of states at the Fermi level. This enhanced DOS at the surface unambiguously confirms the presence of unique surface-localized states that contribute to conduction. Conversely, in the Ru-O slab (Fig. 3(c)), this dispersive projection is heavily diminished near the Fermi level. While some bands with strong intensity persist at higher energies (~ 1 eV), their contribution to electrical transport is negligible due to their thermal inaccessibility and flat dispersion characteristics. This is attributed to the formation of bonds between the surface Ru and O atoms due to oxygen adsorption. As illustrated in the Fig. 3(d), the pristine surface states are suppressed through strong hybridization between the Ru $d$-orbitals and O $p$-orbitals. It is important to note that the bulk DOS of the Ru-O slab (Fig. S3(b)) remains the same as that of the bare Ru slab (Fig. S3(a)), indicating that the significant electronic structure changes are predominantly localized at the surface. This orbital mixing results in marked reduction of the conductive states near the Fermi level compared to the Ru slab. Consequently, this difference signifies the effective suppression of surface states that are essential for conduction, suggesting that the bare Ru slab possesses a greater potential for surface conductivity compared to the oxygen-passivated Ru-O slab.

*Corresponding author.
*E-mail address:* y.lee@inu.ac.kr (Y. Lee).

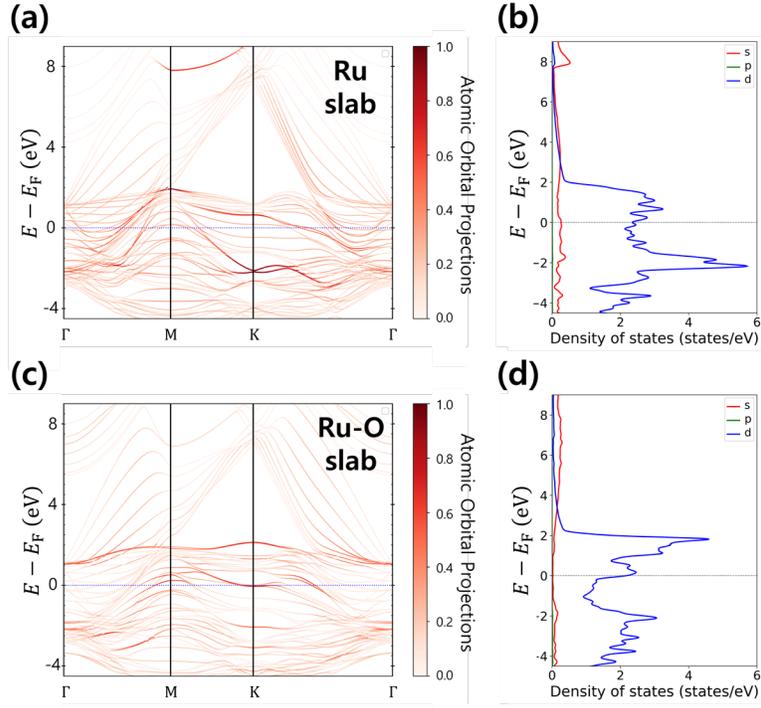

**Fig. 3.** Electronic structure of the 10-layer Ru and Ru-O slabs. Projected band structure of (a) Ru slab and (c) Ru-O slab. The projections are weighted by the orbital contributions from Ru atoms on both surface layers, where the color intensity represents the magnitude of the contribution. Partial density of states projected onto the Ru atoms of both surface layers for the (b) Ru slab and (d) Ru-O slab. The red, green, and blue lines correspond to contributions from Ru $s, p,$ and $d$ orbitals, respectively. The horizontal dotted lines indicate the Fermi level.

To quantify how these electronic structure changes affect macroscopic transport properties, we calculated the electrical resistivity as a function of thickness and compared these properties against bulk Ru resistivity, as shown in Fig. 4. The bulk reference value is the in-plane components ($xx$ and $yy$) of the conductivity tensor for the Ru supercell structure in Fig. 1(a). The thickness values used for resistivity calculations correspond to the distances between the topmost and bottommost atoms. This physical thickness was used to normalize the raw conductivity data from BoltzTraP2, correcting for the normalization, which improperly includes the vacuum region. The Ru slab exhibits a notable behavior as the thickness decreases. In the relatively thick region, the resistivity matches bulk Ru. However, as the metal thickness shrinks, the resistivity decreases. This behavior results from conductive surface states, which provide an additional conduction channel that becomes increasingly dominant as the film becomes thinner. In contrast, the resistivity of the Ru-O slab consistently increases as thickness decreases. This trend is qualitatively similar to experimental observations where oxidation levels lead to an increase in resistivity[54]. While experimental measurements inherently convolve extrinsic factors such as scattering and bulk oxide formation, our calculation isolates the intrinsic electronic structure. Therefore, the resistivity rise observed in our Ru-O slab model reveals that the reduction of surface states is a distinct intrinsic mechanism that modulates transport properties, independent of extrinsic scattering effects. In addition to the absence of the surface states, our calculations inherently capture quantum confinement effects because the slab models are confined in one direction. This leads to a quantization of energy levels and band modulation, distinct from bulk behavior, which effectively reduces the number of available conducting states. While we do not employ a separate quantum confinement model, this reduction in available electron transport pathways due to reduced dimensionality causes a dominant size effect, leading to the observed rise in resistivity as thickness decreases in the absence of surface states[55].

*Corresponding author.
*E-mail address:* y.lee@inu.ac.kr (Y. Lee).

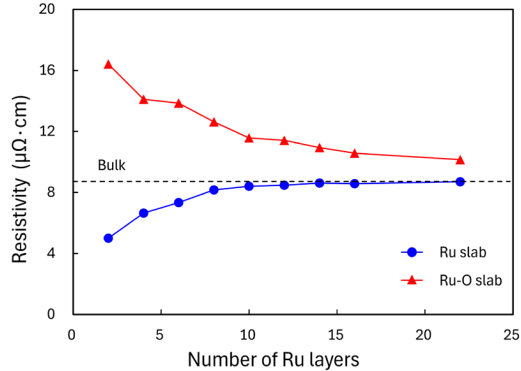

**Fig. 4.** In-plane ($xx$ and $yy$) thickness-normalized resistivities as a function of the number of Ru layers for Ru slabs and Ru-O slabs. The computed resistivity of the bulk Ru is also shown for comparison (horizontal dashed line).

    The direct influence of surface states is quantified in Fig. 5, which compares the conductance ($G$) with the width ($W$) and length ($L$) of the two different slab models. For slabs with a thickness greater than 1.5 nm, the data demonstrates a near-constant offset in conductance between the two models. This stable difference of approximately $4.4 \times 10^{-3}$ S quantifies the intrinsic contribution of the surface states to the total transport. In the ultrathin regime (< 1.5 nm), however, this difference is no longer uniform. This deviation is attributed to the interaction between the two surfaces of the slab. As the film becomes substantially thin, the electronic states on the opposing surface begin to couple with each other. This deforms the surface electronic structure, thereby modifying its transport properties. This approach, which emphasizes the role of intrinsic electronic structure effects, aligns with observation in other metallic systems. For instance, studies on Cu thin films similarly highlight the contribution of surface and interface effects to their thickness-dependent electrical conductivity. Research by N. T. Cuong and S. Okada [31] demonstrated that the total electrical conductivity of bare Cu films can be decomposed into a contribution from Shockley surface states ($\sigma_{Shockley}$) and a thickness-dependent volume contribution. Their work showed that $\sigma_{Shockley}$ provides a substantial and relatively constant contribution, particularly prominent in thinner films, thus acting as an additional conduction channel. Moreover, they observed that orbital hybridization due to the adsorption of oxygen atoms eliminated conductive surface states and significantly reduced electrical conductivity, which is consistent with our Ru-O slab modeling results. Furthermore, a comprehensive study by P. P. Shinde et al. [56] on the thickness-dependent conductivity of Cu thin films under various interfacial conditions provides some insights into how interfaces modulate conductivity. This well-established understanding in Cu, encompassing both intrinsic surface state contributions and the influence of external interfacial environments, underscore the critical role of surface/interface and thickness effects in nanoscopic metallic transport. This provides a robust comparative context for our investigation into Ru's surface state modulation and its implications for next-generation interconnects, reinforcing that similar surface and interface engineering principles will be useful for optimizing performance.

    It is important to note that in this study, the effects of grain boundary (GB) and surface scattering were not considered, while the bulk mean free time was employed in the constant mean free time approximation to model slab resistivity. This simplification was adopted to isolate and highlight the role of surface states and the modulation of the electronic band structure due to quantum confinement. While this approach offers valuable insights into the intrinsic electronic structure effects—as demonstrated in previous studies on copper [31,55,56]—the exclusion of GB and surface scatterings may lead to quantitative deviations from experiments, particularly for ultrathin slabs. However, while the exclusion of scattering may lead to quantitative deviations, the presence of conductive surface states offers an important advantage. As thickness diminishes, these states contribute to conductivity, which acts counter to the resistivity increase from scattering, thereby enabling the film to maintain relatively higher conductivity compared to scenarios where such states are not preserved. Future studies should incorporate these scattering mechanisms to achieve a more comprehensive understanding of electron transport in slab geometries.

*Corresponding author.
*E-mail address:* y.lee@inu.ac.kr (Y. Lee).

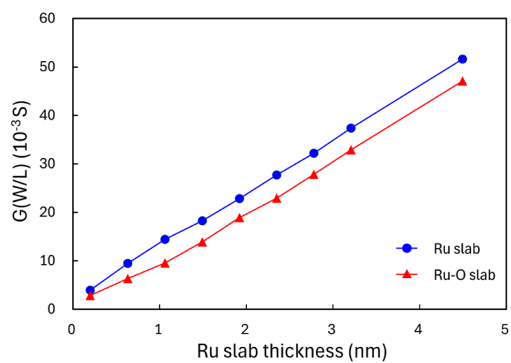

**Fig. 5.** Normalized in-plane conductance as a function of film thickness for Ru slabs with and without surface oxygen atoms. The thickness indicates the distance between the top and bottom Ru atoms.

## 4. Conclusion

This study demonstrates that the electrical transport in ultrathin Ru films is governed not by simple geometric scaling but by their surfaces. We found that bare Ru slabs, which host conductive surface states, exhibit a decrease in resistivity with diminishing thickness, whereas the Ru-O slab shows a monotonic resistivity increase. This opposing behavior identifies the surface electronic structure as the crucial factor in determining charge transport at the nanoscale. From a technological perspective, our results suggest that interface engineering, such as optimizing underlayers or employing appropriate capping layers to mitigate oxidation and preserve conductive surface layers, can provide a practical strategy for reducing resistance in future ultrathin interconnects[56–59].

**Author contributions**

**Gyungho Maeng:** Data curation, Formal analysis, Investigation, Methodology, Software, Validation, Visualization, Writing – original draft. **Subeen Lim:** Formal analysis, Methodology, **Mi Gyoung Lee:** Funding acquisition, Writing – review & editing, **Bonggeun Shong:** Funding acquisition, Validation, Writing – review & editing, **Kyeongjae Cho:** Conceptualization, Validation, Writing – review & editing, **Yeonghun Lee:** Supervision, Conceptualization, Project administration, Formal analysis, Funding acquisition, Methodology, Validation, Writing – review & editing.

**Declaration of competing interest**

The authors declare that they have no known competing financial interests or personal relationships that could have appeared to influence the work reported in this paper.

**Acknowledgments**

This work was supported by Research Project Support Program for Excellence Institute (2024) in Incheon National University. This work was supported by the Technology Innovation Program [Public-private joint investment semiconductor R&D program (K-CHIPS) to foster high-quality human resources] [No. RS-2023-00236667, High performance Ru-TiN interconnects via high temperature atomic layer deposition (ALD) and development on new interconnect materials based on ALD] funded by the Ministry of Trade, Industry & Energy (MOTIE, Korea) (No. 1415187401).

**Data availability**

Data will be made available on request.

*Corresponding author.
*E-mail address:* y.lee@inu.ac.kr (Y. Lee).

*Corresponding author.
*E-mail address:* y.lee@inu.ac.kr (Y. Lee).

*Corresponding author.
*E-mail address:* y.lee@inu.ac.kr (Y. Lee).

*Corresponding author.
*E-mail address:* y.lee@inu.ac.kr (Y. Lee).

*Corresponding author.
*E-mail address:* y.lee@inu.ac.kr (Y. Lee).

# Supplementary Information for
# Role of surface states and band modulations in ultrathin ruthenium interconnects

Gyungho Maeng [a], Subeen Lim [a], Mi Gyoung Lee [b,c], Bonggeun Shong [d], Kyeongjae Cho [e], Yeonghun Lee [a,c,*]

[a]Department of Electronics Engineering, Incheon National University, Yeonsu-gu, Incheon 22012, Republic of Korea

[b]Department of Materials Science and Engineering, Incheon National University, Incheon 22012, Republic of Korea

[c]Research Institute for Engineering and Technology, Incheon National University, Yeonsu-gu, Incheon 22012, Republic of Korea

[d]Major in Advanced Materials and Semiconductor Engineering, Hanyang University, Ansan-si, Gyeonggi-do 15588, South Korea

[e]Department of Materials Science and Engineering, The University of Texas at Dallas, Richardson, TX 75080, USA

[*] Corresponding author.
E-mail address: y.lee@inu.ac.kr (Y. Lee).

# 1. k-points sampling convergence test

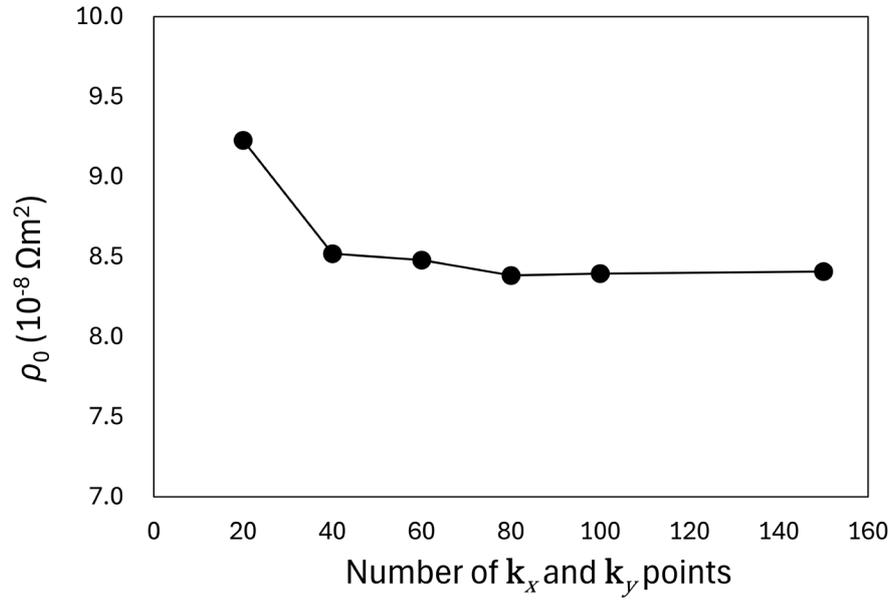

**Fig. S1.** Convergence test of non-self-consistent field calculation with respect to the number of **k**-points in each **k**$_x$ and **k**$_y$ directions

For the electron transport property calculations of the slab model, a convergence test was performed to determine the appropriate **k**-point sampling. Based on this test, a **k**-point grid of 150 × 150 × 6 was used for all subsequent calculations.

## 2. Slab structural information

### 2.1. Atomic structure of ruthenium slab

The cell parameters and atomic positions for the slab model, consisting of 10 atomic layers of Ru, are as follows.

```
CELL_PARAMETERS (angstrom)
   2.7176385168   -0.0000000000   -0.0000000000
  -1.3588192587    2.3535439938   -0.0000000000
   0.0000000000    0.0000000000   36.4496682610

ATOMIC_POSITIONS (angstrom)
  Ru   -0.0000013590    1.5690301137    1.1203561221
  Ru    1.3588206171    0.7845138801    3.1841624504
  Ru   -0.0000013590    1.5690301137    5.3371066920
  Ru    1.3588206171    0.7845138801    7.5073809795
  Ru   -0.0000013590    1.5690301137    9.6528795635
  Ru    1.3588206171    0.7845138801   11.7967913079
  Ru   -0.0000013590    1.5690301137   13.9422914214
  Ru    1.3588206171    0.7845138801   16.1125718607
  Ru   -0.0000013590    1.5690301137   18.2654886488
  Ru    1.3588206171    0.7845138801   20.3293122591
```

### 2.2. Atomic structure of oxygen-terminated slab

The cell parameters and atomic positions for the slab model, featuring oxygen atoms adsorbed on both surfaces of the 10-layer Ru slab, are as follows.

```
CELL_PARAMETERS (angstrom)
   2.7176385168    0.0000000000    0.0000000000
  -1.3588192587    2.3535439938    0.0000000000
   0.0000000000    0.0000000000   40.7396019132

ATOMIC_POSITIONS (angstrom)
 O    1.3588206170    0.7845138800    1.9215263348
 Ru  -0.0000013590    1.5690301140    3.1267286058
 Ru   1.3588206170    0.7845138800    5.3525425174
 Ru  -0.0000013590    1.5690301140    7.5021752398
 Ru   1.3588206170    0.7845138800    9.6383902630
 Ru  -0.0000013590    1.5690301140   11.7943624494
 Ru   1.3588206170    0.7845138800   13.9446348411
 Ru  -0.0000013590    1.5690301140   16.1007213215
 Ru   1.3588206170    0.7845138800   18.2371460618
 Ru  -0.0000013590    1.5690301140   20.3871107695
 Ru   1.3588206170    0.7845138800   22.6134848676
 O   -0.0000013590    1.5690301140   23.8187882072
```

## 3. Local potential distribution

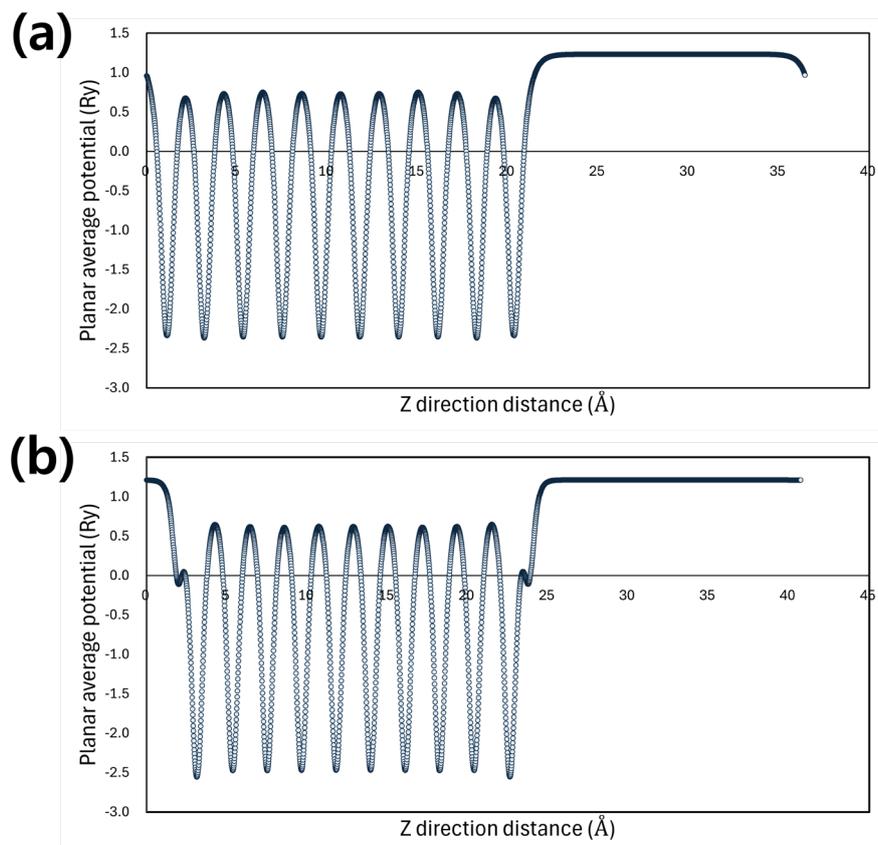

**Fig. S2.** Planar-averaged electrostatic potential along the Z-direction for (a) the Ru slab and (b) Ru-O slab.

The potential profiles exhibit flat behavior in whole regions and are symmetric along the slab normal, confirming the absence of any macroscopic electric field or built-in potential across the entire slab. This analysis ensures that our calculations are free from spurious electric field artifacts.

## 3. Layer-resolved projected density of states

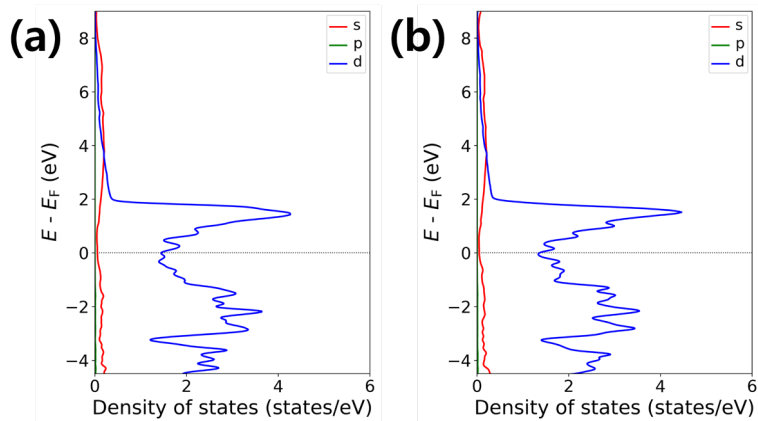

**Fig. S3.** Bulk density of states (DOS) for (a) Ru slab and (b) Ru-O slab. The red, green, and blue lines represent the contributions from Ru *s*, *p*, and *d* orbitals, respectively. The dotted line indicates Fermi level.

A comparison between Fig. S(a) and Fig. S(b) shows that the bulk electronic structure of both Ru and Ru-O slabs is quite similar. Thus, we assume that the significant changes observed in the electronic structure upon oxygen adsorption are predominantly localized at the surface.